\documentclass[prd,twocolumn,superscriptaddress,showpacs,preprintnumbers,amsmath,amssymb]{revtex4}

\usepackage{graphicx}
\usepackage{dcolumn}
\usepackage{bm}

\begin{document}

\preprint{}

\title{Self-Organized Criticality in a Spherically Closed Cellular Automaton: \\ Modeling Soft Gamma Repeater Bursts Driven by Magnetic Reconnection}

\author{Ken'ichiro Nakazato}
 \email{nakazato@rs.tus.ac.jp}
 \affiliation{Department of Physics, Faculty of Science \& Technology, Tokyo University of Science, 2641 Yamazaki, Noda, Chiba 278-8510, Japan
}%

\date{\today}

\begin{abstract}
A new cellular automaton (CA) model is presented for the self-organized criticality (SOC) in recurrent bursts of soft gamma repeaters (SGRs), which are interpreted as avalanches of reconnection in the magnetosphere of neutron stars. The nodes of a regular dodecahedron and a truncated icosahedron are adopted as spherically closed grids enclosing a neutron star. It is found that the system enters the SOC state if there are sites where the expectation value of the added perturbation is nonzero. The energy distributions of SOC avalanches in CA simulations are described by a power law with a cutoff, which is consistent with the observations of SGR 1806$-$20 and SGR 1900+14. The power-law index is not universal and depends on the amplitude of the perturbation. This result shows that the SOC of SGRs can be illustrated not only by the crust quake model but also by the magnetic reconnection model.
\end{abstract}


\maketitle

\section{Introduction} \label{intro}
Self-organized criticality (SOC) proposed by Bak~{\it et al.} \cite{bak87,bak88} has revealed a wide range of mechanisms of complex behavior occurring in nature. According to the concept of SOC, a nonequilibrium open system evolves spontaneously into a critical state characterized by a power-law distribution of its physical quantity. One of the most well-known examples of an SOC system is earthquakes, where the energy supplied by plate motion is dissipated in the crust. The SOC of earthquakes has been illustrated with a sand-pile model, and an empirical power-law relation between the size and frequency of seismic events has been demonstrated \cite{bak89,ofc92}; a related model for the propagation of brittle failure had previously shown power laws \cite{katz86}.

SOC models fit some properties of astrophysical phenomena, for instance, solar flares \cite{asch14}. The recurrent bursts of soft gamma repeaters (SGRs) have event energy distributions well fitted with a power law \cite{gog99,gog00,prka12}. It is currently thought that SGRs are associated with ultrastrongly magnetized neutron stars ($\gtrsim$10$^{14}$~G). They generally undergo the recurrent emission of soft gamma rays with a short duration ($\sim$0.1~s). These bursts have been suggested to be due to neutron star crust fractures driven by the stress of an evolving magnetic field \cite{thdu95}, which are called starquakes. Recently, however, Link~\cite{link13} pointed out that it is difficult to reproduce the typically observed rise time of emission ($\lesssim$10~ms) if the energy is deposited deep in the crust or deeper. The trapped seismic energy takes seconds to minutes to reach the stellar surface and cause burst emissions. As a corollary, the energy should be released not inside the star but in the magnetosphere \cite{katz82,lyut03,lyut06,kbl07,gillheyl10}.

In this paper, we present a new cellular automaton (CA) model that mimics SOC avalanches in SGRs. We first follow the CA model for solar flares proposed by Lu and Hamilton \cite{luha91}, which is expressed as discretized magnetohydrodynamic (MHD) equations \cite{isli98}. In their model, solar flares are interpreted as avalanches of many small magnetic reconnections. Similarly, we assume that the energy is released in the magnetosphere of neutron stars through MHD instabilities, such as the tearing mode \cite{lyut03,kbl07}. In the original model by Lu and Hamilton, a three-dimensional Cartesian coordinate grid of points is used. However, a neutron star may be sufficiently small for the size of avalanches to be comparable to the system size. In fact, since the energy of a magnetic field $B$ in a region of volume $L^3$ is $E=B^2 L^3/8\pi$, the length scale is $L\sim 10^4$~cm with $B=10^{14}$~G and $E=10^{38}$~erg, which is the {\it lowest} observed energy of short bursts \cite{gog99,gog00}. Note that the radius of neutron stars is typically $R\sim10^6$~cm. Furthermore, the perturbation rule in the original model by Lu and Hamilton tends to increase the magnetic field in a certain direction. Here, we use the grid with a closed geometry and modify the perturbation rule to satisfy a conservation law for magnetic field.

\section{Cellular automaton model} \label{camodel}
\subsection{grids} \label{grids}
We introduce ``spherically'' {\it closed} grids for our CA model to map the surface enclosing a neutron star. To avoid grid anisotropy, we require that (i) the shape is nearly spherical, (ii) all edges are of equal length and (iii) all nodes are equivalent. Among the polyhedra which satisfy these conditions, the nodes of a regular dodecahedron and a truncated icosahedron (the shapes of a soccer ball and fullerene C$_{60}$), which have $n=20$ and 60 nodes, respectively, are adopted as our grids. They have a closed geometry, in marked contrast to CA models for SOC systems studied so far. Incidentally, Schein and Gayed \cite{schgay14} recently presented the techniques to construct the icosahedral Goldberg polyhedra, which are nearly spherical grids with more nodes. These grids satisfy the conditions (i) and (ii) but not (iii). Although the investigation with more nodes is interesting, we defer it to future work.

In the CA simulation, we assign the values of $B_i$ for the $i$-th site of the grids, which represent the deviation of the radial component of the magnetic field from the unperturbed background configuration. We set $B_i=0$ for all sites at the initial time.

\subsection{step 1: perturbations} \label{step1}
As the first step of our CA rule, we add perturbations to $B_i$. This perturbation corresponds to the bending of magnetic field line outward because $B_i$ is the deviation of the radial component. Since the CA simulation is carried out on the closed surface ($S$) in our model, the magnetic field should be perturbed to satisfy a conservation law,
\begin{equation}
\oint_S \bm{B} \cdot \bm{n} \, {\rm d}S = 0 \quad \Longrightarrow \quad \sum_{i=1}^{n} B_i = 0,
\label{magcons}
\end{equation}
where $\bm{n}$ is the outward-pointing unit normal vector of the surface $S$. Therefore, we choose two neighboring sites and add a positive perturbation ($+\Delta B$) to one and a negative perturbation ($-\Delta B$) to the other. The perturbation, $\Delta B$, is given as
\begin{equation}
\Delta B = \langle \Delta B \rangle | \sigma |,
\label{gaupert}
\end{equation}
where $\sigma$ is a random number obeying a Gaussian distribution with zero mean and unit variance. The amplitude of the perturbation, $\langle \Delta B \rangle$, is a model parameter as discussed later. 

\begin{figure}
\begin{center}
\includegraphics[height=.33\textheight]{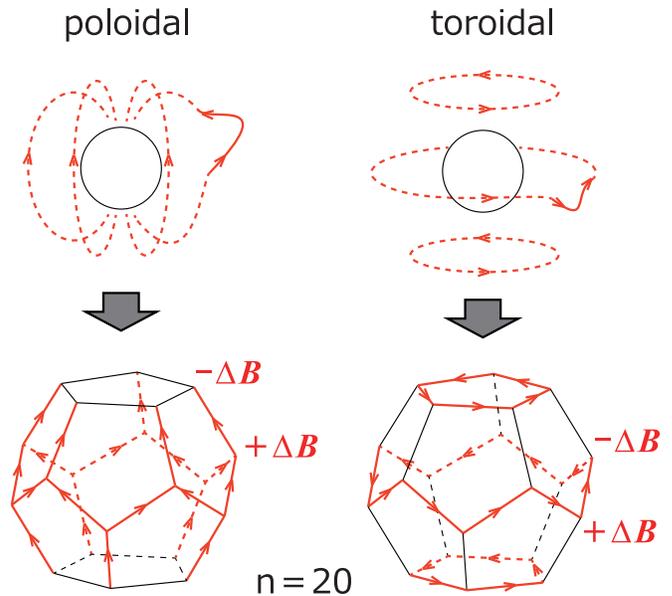}
\caption{Schematic diagrams of poloidal and toroidal magnetic field lines in the magnetosphere of neutron stars and their mapping on a regular dodecahedron ($n=20$) grid. Positive ($+\Delta B$) and negative ($-\Delta B$) perturbations are added at the start point and end point of the edges shown as thick lines, respectively.}
\label{procd}
\end{center}
\end{figure}

The sites where the perturbation is added are chosen as follows. In our model, we assume the bending of magnetic field line outward similarly to the rise of the magnetic field line on the surface of the Sun. Then positive and negative perturbations appear at the footpoints of the magnetic field line. For simplicity, we deal with two cases for the unperturbed background configuration, namely, poloidal and toroidal magnetic fields, and randomly choose one of the edges considered to be along the background field lines. The perturbations are added to the sites corresponding to both ends and the polarity of the magnetic field line is reflected in the sign of the perturbations. This procedure is illustrated in Figure~\ref{procd}.

The edges along the background field lines are determined as below. At first, for poloidal case, we choose edges connecting nodes with the different latitude. The number of such edges is 20 for a regular dodecahedron with 30 edges and 60 for a truncated icosahedron with 90 edges. Then the polarity is assigned for the end points of each edge reflecting their latitudes. Next, for toroidal case, we choose the edges to construct the circuits. In this process, the edges connecting nodes with the same longitude are not chosen and the number of edges considered is set to be same with the poloidal case. For a regular dodecahedron, circuits with 5 edges, 10 edges and 5 edges are constructed as shown in the bottom right of Figure~\ref{procd}. For a truncated icosahedron, circuits with 5 edges, 15 edges, 20 edges, 15 edges and 5 edges are constructed.

\subsection{step 2: reconnections} \label{step2}
The second step of our CA rule is reconnection. We define the magnetic field stress, which is the difference between the local magnetic field and the average of its three nearest neighbors $B_{\rm nn}$, as
\begin{equation}
{\rm d} B_i = B_i -\frac{1}{3}\sum_{\rm nn} B_{\rm nn}.
\label{defdb}
\end{equation}
The site is unstable to reconnection when the absolute value of its magnetic stress exceeds some critical value $B_{\rm c}$ \cite{luha91,dimit11},
\begin{equation}
| {\rm d} B_i | > B_{\rm c}.
\label{reccond}
\end{equation}
This criterion is reasonable because a large magnetic stresses favors magnetic reconnection \cite{dimit11}. If a reconnection instability occurs, the magnetic field stress is canceled as follows:
\begin{equation}
B_i \to B_i - \frac{3}{4} {\rm d} B_i, \qquad B_{\rm nn} \to B_{\rm nn} + \frac{1}{4} {\rm d} B_i,
\label{carule}
\end{equation}
resulting in ${\rm d} B_i \to 0$. If the nearby sites become unstable owing to the reconfiguration expressed by (\ref{carule}), additional reconnection events occur, leading to a large avalanche in some cases.

\subsection{step 3: event energy} \label{step3}
When all instabilities in the grid have been relaxed, we return to the first step of adding perturbations. The energy released during the avalanche is defined as
\begin{equation}
E = \sum_{i=1}^n \left|B_{i}^{(0)}\right|^2 - \sum_{i=1}^n \left|B_{i}^{(1)}\right|^2,
\label{enerls}
\end{equation}
where the superscripts (0) and (1) denote the values before and after the avalanche, respectively.

Note the dimensionless quantities in our CA model. The threshold for the reconnection instability is fixed at $B_{\rm c}=4$ in this study. Since the cases with same values of $\langle \Delta B \rangle / B_{\rm c}$ are equivalent, we only investigate the dependence on $\langle \Delta B \rangle$. If a reconnection event occurs at a single site $i$, the released energy is $\frac{3}{4}|{\rm d} B_i|^2$. Therefore, the minimum released energy in our CA model is $E_{\min} =12$ for $B_{\rm c}=4$.

\section{Results} \label{results}
\subsection{poloidal versus toroidal} \label{polotoro}
\begin{figure}
\begin{center}
\includegraphics[scale=0.84]{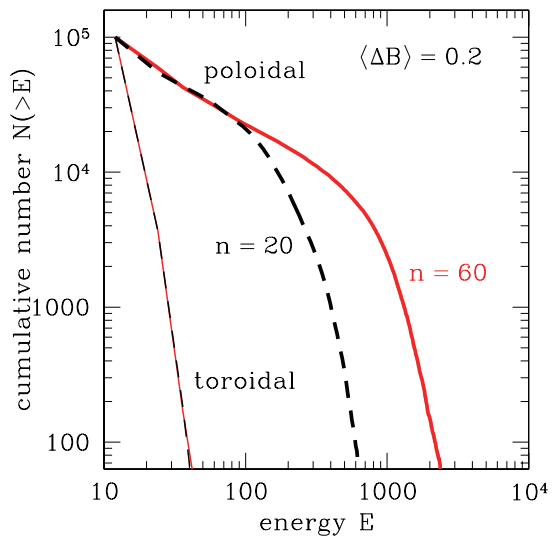}
\caption{Cumulative burst energy distributions, $N(>\!\!E)$, for the case of $\langle \Delta B \rangle = 0.2$. Thick and thin lines represent the models with poloidal and toroidal perturbations, respectively. Dashed and solid lines represent the models with the regular dodecahedron ($n=20$) grid and truncated icosahedron ($n=60$) grid, respectively.}
\label{rslt1}
\end{center}
\end{figure}

Now we move on to the results of our CA simulations. In Figure~\ref{rslt1}, cumulative energy distributions of avalanches, $N(>\!\!E)$, are plotted for the case of $\langle \Delta B \rangle = 0.2$. As can be seen, the models with perturbations under poloidal and toroidal fields have considerably different profiles. In the models with poloidal perturbations, large avalanches with the size of the system occur, which is one of the characteristic features of SOC. In contrast, the models with toroidal perturbations do not reach the SOC state. There are very few bursts with energy $\gtrsim \! 3E_{\min}$ while small bursts occur accidentally. In the toroidal models, the {\it time} average of the random perturbations is zero at each site (see Figure~\ref{procd}). This result is consistent with that of Lu and Hamilton \cite{luha91}, where SOC was not found with the random perturbation being symmetric about zero. However, in the poloidal models, the system reaches the critical state while the {\it spatial} average of the perturbations is zero. In this case, the time average of each node depends on the latitude. It is positive for one hemisphere but negative for the other hemisphere. This is because, as already stated, the polarity is assigned reflecting the latitude.

\begin{figure}
\begin{center}
\includegraphics[scale=0.84]{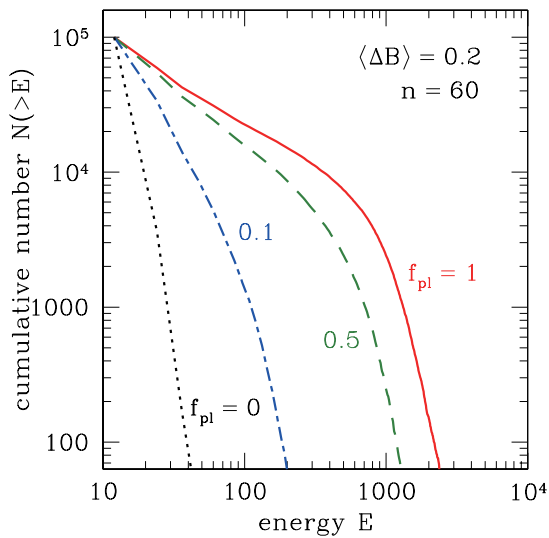}
\caption{Same as Figure~\ref{rslt1} but for the models with the truncated icosahedron ($n=60$) grid. Solid, dashed, dot-dashed and dotted lines correspond to the cases with $f_{\rm pl}=1$ (purely poloidal), 0.5, 0.1 and 0 (purely toroidal), respectively.}
\label{rslta}
\end{center}
\end{figure}

Here, we investigate the mixed case of poloidal and toroidal fields. In the step 1 of the CA rule, we choose the poloidal perturbation or toroidal perturbation randomly and the probability of poloidal is set to $f_{\rm pl}$. Thus the purely poloidal and toroidal cases correspond to $f_{\rm pl}=1$ and $f_{\rm pl}=0$, respectively. The dependence of $f_{\rm pl}$ is shown in Figure~\ref{rslta} for the cases of $\langle \Delta B \rangle = 0.2$ and the truncated icosahedron model with $n=60$ nodes. As already mentioned, in our CA model, bursts with $\lesssim \! 3E_{\min}$ happen even if the system does not reach the SOC state. Thus the distributions with $E \lesssim 3E_{\min}$ may include a contamination and we pay attention to the events with $\gtrsim \! 3E_{\min}$. For the case with $f_{\rm pl}=0.5$, where the poloidal and toroidal fields are regarded as comparable, large avalanches are still observed. On the other hand, for the dominantly toroidal case ($f_{\rm pl}=0.1$), it is hard to find the SOC feature.

\subsection{parameter dependence} \label{paredpd}
Hereafter, we focus on the purely poloidal models. The cumulative distributions appear to decrease rapidly at high energies owing to finite-size effects. In fact, the cutoff energy of the truncated icosahedron model with $n=60$ nodes is higher than that of the regular dodecahedron model with $n=20$. A cutoff feature was also found in the observations of SGR 1806$-$20 and SGR 1900+14 \cite{prka12}. While the statistical significance of the observed cutoff is not high, bursts with the highest energy may correspond to avalanches enclosing the central neutron star. On the other hand, the distributions of the truncated icosahedron model and regular dodecahedron model are similar at low energies, where finite-size effects are negligible.

\begin{figure}
\begin{center}
\includegraphics[scale=0.84]{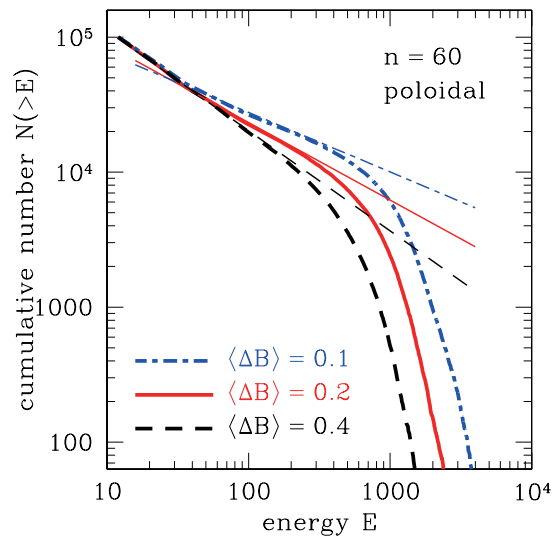}
\caption{Same as Figure~\ref{rslt1} but for the models with the truncated icosahedron ($n=60$) grid and poloidal perturbations. Thick dot-dashed, thick solid and thick dashed lines correspond to the cases with $\langle \Delta B \rangle = 0.1$, 0.2 and 0.4, respectively. Thin lines are power laws $N(> \!\! E) \propto E^{1-\gamma}$ with $\gamma=1.44$ (dot-dashed), 1.57 (solid) and 1.74 (dashed).}
\label{rslt2}
\end{center}
\end{figure}

We show the dependence on the amplitude of perturbation $\langle \Delta B \rangle$ in Figure~\ref{rslt2}, where the cumulative distributions are plotted for the models with the truncated icosahedron ($n=60$) grid and poloidal perturbations. We find that the distributions with $E \gtrsim 3E_{\min}$ are well fitted with power laws. The power-law index, $\gamma$, of the differential distributions ${\rm d}N \propto E^{-\gamma} {\rm d}E$ is related to the power-law fit of the cumulative distributions as $N(> \!\! E) \propto E^{1-\gamma}$. According to the observations of SGR 1806$-$20 and SGR 1900+14, the energy distributions of bursts have been fitted with $\gamma=1.43$-1.76 \cite{gog99,gog00} and $\gamma \sim 1.55$ \cite{prka12}. In our CA simulation, $\gamma$ depends on the value of $\langle \Delta B \rangle$ and it is 1.44, 1.57 and 1.74 for $\langle \Delta B \rangle = 0.1$, 0.2 and 0.4, respectively. Here, Kolmogorov-Smirnov probabilities of the fitting in the range of $3E_{\min} < E < 100E_{\min}$ are $>$0.999, $>$0.999 and 0.964 for $\langle \Delta B \rangle = 0.1$, 0.2 and 0.4, respectively. On the other hand, in the range of $E > 3E_{\min}$, Kolmogorov-Smirnov probabilities are 0.000 for all models due to the cutoff.

\begin{figure}
\begin{center}
\includegraphics[scale=0.84]{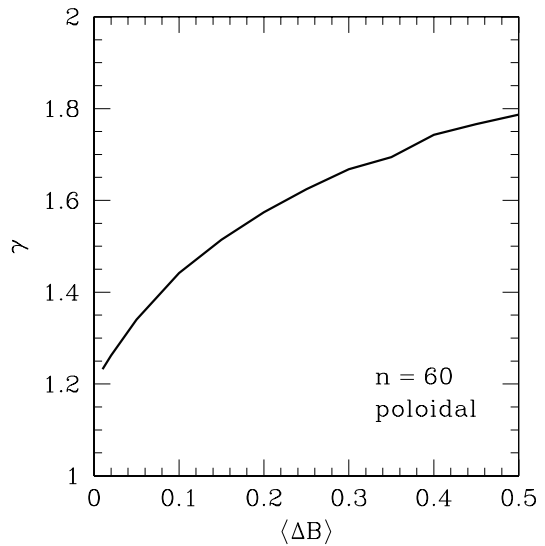}
\caption{Power-law index $\gamma$ as a function of the amplitude of perturbation $\langle \Delta B \rangle$ for the models with the truncated icosahedron ($n=60$) grid and poloidal perturbations.}
\label{rslt3}
\end{center}
\end{figure}

From Figure~\ref{rslt3}, we can see that the power-law index $\gamma$ increases with the amplitude of perturbation $\langle \Delta B \rangle$. The most natural interpretation for this is that, for large $\langle \Delta B \rangle$, the perturbation of one site can grow exclusively and the site tends to become unstable before accumulating perturbations of other sites. As a result, the number of large avalanches, i.e. with a large energy, is reduced. The lack of universality of the power-law index is an interesting feature that was also observed in the model of Olami~{\it et al.} \cite{ofc92}.

\section{Conclusion and discussion} \label{concdis}
In conclusion, we have shown the SOC behavior of a new CA model with spherically closed grids so as to demonstrate the recurrent bursts of SGRs. We adopted the nodes of a regular dodecahedron and a truncated icosahedron as our CA grids. Perturbations were added to satisfy a conservation law (\ref{magcons}) and the sign was determined from the polarity of the unperturbed background magnetic field line. For the configuration of the unperturbed field, both poloidal and toroidal cases were considered. We found that the SOC state is reached only for the poloidal case owing to the existence of sites where the expectation value of the added perturbation is nonzero.

The cumulative burst energy distribution has a cutoff at high energies owing to finite-size effects. Similarly, the cutoff found in the observations of SGRs may be attributed to the compactness of the magnetosphere of neutron stars. In addition to the recurrent bursts, giant flares with enormous energy ($\sim$10$^{44}$-10$^{47}$~erg) and long duration ($\sim$100~s) are occasionally observed from SGRs \cite{maze79,hurl99,hurl05,palm05,tera05}. The energy release in the magnetosphere of neutron stars was proposed early on as an explanation of a giant flare \cite{katz82}. Taking the large difference in emission energies into account, it is difficult for our CA model to describe the giant flares, which might be produced by global field reconfiguration \cite{masa10}.

The power-law index, $\gamma$, found in our CA model depends on the amplitude of perturbation, $\langle \Delta B \rangle$, and ranges from 1.2 to 1.8. This value is consistent with the observations. According to Prieskorn and Kaaret \cite{prka12}, there is no difference between the power-law indices observed in high- and low-burst-rate regimes. This result may appear to contradict our CA model with $\langle \Delta B \rangle$ dependence, but we found that cases with the same values of $\langle \Delta B \rangle / B_{\rm c}$ are equivalent. If the amplitude of perturbation and the criterion for reconnection depend on the unperturbed background magnetic field likewise (for instance, they are linearly related), $\langle \Delta B \rangle / B_{\rm c}$ and $\gamma$ may be almost insensitive to the environment. In any case, further investigation of the relation between the power-law index and burst rate is important.

Comparing with the original model by Lu and Hamilton \cite{luha91}, the grid and the perturbation rule for magnetic field are modified in this study. On the other hand, the CA rule of reconnection is identical with that in Lu and Hamilton model while the number of nearest neighbors is different. This CA model relates to the induction equation,
\begin{equation}
\frac{\partial \bm{B}}{\partial t} = \bm{\nabla} \times (\bm{V} \times \bm{B}) + \eta \nabla^2 \bm{B},
\label{induc}
\end{equation}
where $\bm{V}$ and $\eta$ are the plasma velocity and the resistivity, respectively \cite{isli98,dimit11}. Note that the Laplacian of the magnetic field $\nabla^2 \bm{B}$ corresponds to the right hand side of equation (\ref{defdb}) while the sign is opposite. Thus, if the reconnection occurs, the variation of $B_i$ is proportional to $-{\rm d} B_i$ as in equation (\ref{carule}). On the other hand, the perturbation of this CA model is interpreted as the convective term $\bm{\nabla} \times (\bm{V} \times \bm{B})$. As already mentioned, this CA model was originally proposed for solar flares. In contrast, we have applied it to recurrent bursts of SGRs. The modifications for SGRs, such as the relativistic forms of the MHD equations and QED effects, will be interesting issues.

In this paper, we have shown that the SOC of SGRs can be illustrated not only by the crust quake model but also by the magnetic reconnection model. To mimic magnetic reconnections, other CA rules have been examined in the context of solar flares \cite{lu93,geo95,vass98,mocha10}. Incidentally, we have revised the CA rule for the redistribution from equation (\ref{carule}) to that proposed by Lu~{\it et al.} \cite{lu93}, but the qualitative features remained unchanged. It would also be interesting to study the distributions of other SOC parameters (avalanche size, waiting time and so forth) \cite{asch14,mcin02}, which will be presented elsewhere.

\begin{acknowledgments}
The author is grateful to Masanao Sumiyoshi for useful discussions and valuable comments. This work was partially supported by Grants-in-Aids for the Scientific Research (Nos.~24105008, 26870615) from MEXT in Japan.
\end{acknowledgments}

\bibliographystyle{apsrev}
\bibliography{apssamp}

\end{document}